\documentstyle[12pt,aaspp4]{article}
\def\tempest%
{\begin{array}{ccc}
1 & 1 & 1 \\
1 & 1 & 1 \\
4 & 3 & 8
\end{array}}

\def\kms{{\rm km}\,{\rm s}^{-1}}
\def\kpc{{\rm kpc}}
\def\dls{{D_{\rm LS}}}
\def\te{{t_{\rm E}}}
\def\re{{r_{\rm E}}}

\begin{document}

\title{The Relative Lens-Source Proper Motion in MACHO 98-SMC-1}

\author{
M. D. Albrow\altaffilmark{1}, 
J.-P. Beaulieu\altaffilmark{2},
J. A. R. Caldwell\altaffilmark{3}, 
D. L. DePoy\altaffilmark{4}, 
M. Dominik\altaffilmark{2}, 
B. S. Gaudi\altaffilmark{4}, 
A. Gould\altaffilmark{4}, 
J. Greenhill\altaffilmark{5}, 
K. Hill\altaffilmark{5},\\
S. Kane\altaffilmark{5,6}, 
R. Martin\altaffilmark{7},  
J. Menzies\altaffilmark{3}, 
R. M. Naber\altaffilmark{2}, 
K. R. Pollard\altaffilmark{1},\\ 
P. D. Sackett\altaffilmark{2}, 
K. C. Sahu\altaffilmark{6}, 
P. Vermaak\altaffilmark{3},  
R. Watson\altaffilmark{5}, 
A. Williams\altaffilmark{7}
}
\author{The PLANET Collaboration}
\centerline{and}
\centerline{R. W. Pogge\altaffilmark{4}}
\affil{}

\altaffiltext{1}{Univ. of Canterbury, Dept. of Physics \& Astronomy, 
Private Bag 4800, Christchurch, New Zealand}
\altaffiltext{2}{Kapteyn Astronomical Institute, Postbus 800, 
9700 AV Groningen, The Netherlands}
\altaffiltext{3}{South African Astronomical Observatory, P.O. Box 9, 
Observatory 7935, South Africa}
\altaffiltext{4}{Ohio State University, Department of Astronomy, Columbus, 
OH 43210, U.S.A.}
\altaffiltext{5}{Univ. of Tasmania, Physics Dept., G.P.O. 252C, 
Hobart, Tasmania~~7001, Australia}
\altaffiltext{6}{Space Telescope Science Institute, 3700 San Martin Drive, 
Baltimore, MD. 21218~~U.S.A.}
\altaffiltext{7}{Perth Observatory, Walnut Road, Bickley, Perth~~6076, Australia}


\begin{abstract}
 
We present photometric and spectroscopic data for the second microlensing
event seen toward the Small Magellanic Cloud (SMC), MACHO-98-SMC-1.
The lens is a binary.  We resolve the
caustic crossing and find that the source took $2\Delta t = 8.5$ hours to
transit the caustic.  We measure the
source temperature $T_{\rm eff}=8000$~K both spectroscopically and from the
color $(V-I)_0\sim 0.22$.  We find two acceptable binary-lens models.
In the first, the source crosses the caustic at $\phi=43^\circ\hskip-2pt .2$
and the unmagnified source magnitude is $I_s=22.15$.  
The angle implies that the lens
crosses the source radius in time $t_*=\Delta t \sin\phi = 2.92$ hours.
The magnitude (together with the temperature) implies that the angular
radius of the source is $\theta_* = 0.089\,\mu$as.  Hence, the proper motion is
$\mu=\theta_*/t_*=1.26\,\kms\,\kpc^{-1}$.  For the second solution, the
corresponding parameters are
$\phi=30^\circ\hskip-2pt .6$, $I_s=21.81$, $t_*=2.15$ hours, 
$\theta_* = 0.104\,\mu$as, $\mu=\theta_*/t_*=2.00\,\kms\,\kpc^{-1}$.
Both proper-motion estimates are slower than 99.5\% of the proper motions 
expected for halo lenses.  Both are consistent with an ordinary binary lens 
moving at $\sim 75$--$120\,\kms$ within the SMC itself.  
We conclude that the lens is most likely in the SMC proper.

\end{abstract}


\newpage
\section{Introduction}

When microlensing searches toward the Large Magellanic Cloud (LMC) were 
initiated seven years ago, it was believed that the microlensing events
themselves could resolve the question of
whether the Galactic dark halo is composed primarily of baryonic
material in the form of Massive Compact 
Halo Objects (MACHOs).  The major potential difficulty was thought to be 
the problem of distinguishing lensing events from variable stars.

The situation today differs markedly from these expectations.  On the one 
hand, more than a dozen candidate events have been detected toward the LMC 
(Aubourg et al.\ 1993; Alcock et al.\ 1997a), and while a few may be 
intrinsic variables,
the great majority are clearly due to genuine microlensing events.  On the 
other hand, the interpretation of these events is far from clear.  The 
optical depth is $\tau \sim 2.5 \times 10^{-7}$ (Alcock et al.\ 1997a), 
about half what would be 
expected from a full dark 
halo of MACHOs, and far more than could be due to known 
populations of stars along the line of sight in the Galactic disk, Galactic 
spheroid, and LMC disk.  By contrast, from the 
measured Einstein crossing 
times $\te \sim 45~{\rm{days}}$ and the well-constrained kinematics of the 
halo, the mass of these objects would appear to be $0.4\, M_\odot$.  These 
could not be made of hydrogen or the population would be discovered easily.
The paucity of obvious solutions has led to a large number of 
alternative explanations for the observed events, including stars within the
LMC bar/disk (Sahu 1994), a 
disrupted dwarf galaxy in front of the LMC (Zhao 1998, Zaritsky \& Lin 
1997), a warped and flared Milky Way disk (Evans et al.\ 1998), 
non-standard halo kinematics, 
and a halo composed of white dwarfs.  However, all of these solutions present
considerable difficulties 
(Gould 1995, 1998a; Beaulieu \& Sackett 1998; Bennett 1998; 
Fields, Freese, \& Graff 1998;
Gyuk \& Gates 1998).

More observations of the type already in progress seem unlikely to resolve 
the issue by themselves; new methods and strategies will be required.  
One approach is to search for microlensing toward the Small Magellanic Cloud 
(SMC).  If the LMC events are due to halo lenses, these should produce 
events of similar duration and frequency, though not identical 
if the halo is significantly flattened (Sackett \& Gould 1993).  MACHO 
(Alcock et al.\ 1997b), EROS (Palanque Delabrouille et al.\ 1998) and OGLE
(Udalski et al.\ 1998) are carrying out such observations but, 
since the SMC has
many fewer stars (and hence fewer expected events), this strategy is unlikely 
to resolve the matter by itself within the next few years.

Another approach is intensive photometric monitoring of ongoing events with 
the aim of measuring either ``parallax'' or ``proper motion.''  These measure 
respectively $v/x$ and $v/(1-x)$ where $v$ is the transverse speed of the 
lens relative to the observer-source line of sight, and $x$ is the ratio of 
the lens to source distance, and can distinguish between halo and LMC 
lenses almost on a case-by-case basis (Boutreux \& Gould 1996).
Typical events are long enough that the Earth's motion would often cause 
detectable ``parallax'' deviations in the light curves if the lenses were 
in the halo, provided that all events were monitored intensively
for their entire duration (Gould 1998b).  GMAN (Alcock et al.\ 1997c) is 
currently following all LMC and SMC events with the aim of detecting this
effect, although probably with insufficient intensity to do so in most cases.
EROS and MACHO have recently jointly proposed much more intensive follow-up
observations at the European Southern Observatory, but have not as yet 
been awarded telescope time.  Proper motions can be measured if a 
lens caustic (region of formally infinite magnification) passes over the 
face of the source.  The finite size of the
source suppresses this infinite magnification according to its size.  By 
modeling the light curve, one can therefore measure 
$t_*$, the time 
it takes the lens to cross the source radius.  
The angular size of the source, 
$\theta_*$ can be estimated from its color, flux and the Planck law.  
The proper motion is then given by,
\begin{equation}
\mu = { \theta_* \over t_*} \label{eqn:mudef}
\end{equation}

Unfortunately, the majority of events are caused by point-mass lenses for 
which the caustic is a single point.  The probability that this point will 
pass over the face of a typical source is $\la 10^{-2}$.  However, for binary 
lenses of near-equal mass ratio, 
the caustics are closed curves whose size is of order the Einstein 
radius.  This means that when there is one crossing, there is usually a 
second -- with forewarning.

On 8 June 1998, MACHO (http://darkstar.astro.washington.edu)
issued an alert for such a second caustic crossing for the second event
discovered toward the SMC, MACHO-98-SMC-1 
(J2000 $\alpha=$00:45:35.2, $\delta=-72$:52:34).  
This alert provided 
a rare opportunity to determine the lens location for at least one SMC event. 

\section{Observations}

The Probing Lensing Anomalies NETwork (PLANET) was formed to monitor 
ongoing microlensing events on a round-the-clock basis from observatories
girding the southern hemisphere with a view to detecting anomalies in the 
light curves which might
betray the presence of a planetary companion to the lensing object 
(Albrow et al.\ 1998).  We have obtained substantial dedicated
telescope time during the bulge season (southern Autumn and Winter) on the 
SAAO 1 m (South Africa, $20^\circ$ $49'$, $-32^\circ$), the CTIO-Yale 1 m 
(Chile, $289^\circ$ $11'$, $-30^\circ$), the Canopus 1 m 
(Tasmania, $147^\circ$ $32'$, $-43^\circ$), and the Perth 0.6 m 
(Western Australia, 
$116^\circ$ $8'$, $-32^\circ$).  In addition, we fortuitously obtained 
additional
observations from the CTIO 0.9 m.  All of our previous monitoring was 
of bulge events because they are the most common and are 
concentrated in a relatively compact observing season.  However, in view 
of the importance of MACHO-98-SMC-1, we focused a major effort upon it.

Nightly observations (weather permitting) were made from the SAAO 1 m, the
CTIO-Yale 1 m, the CTIO 0.9 m, and the Canopus 1 m 
from 9 June to 17 June.  These consisted typically of 5 to 20 min 
exposures depending on the source brightness, 
mostly in Cousins $I$, but with some in Johnson $V$.  Reductions were performed
at the telescope as detailed in Albrow et al.\ (1998).
On 15 June, MACHO 
predicted  that the second crossing would occur on $19.2 \pm 1.5$ June UT on
the basis of a binary-lens fit which they put on their web site.
From PLANET observations alone (and confirmed with the addition of
earlier MACHO observations posted on their Web site) we were able on 17 June
to refine the prediction of the peak to be 18.0 June UT and issued our
own caustic alert including light-curve data.  For the next two days, we
devoted all 
available telescope time to this event, and then continued sparser observations
into the beginning of July.

	In addition, we obtained a spectrum of the source during the
caustic crossing on 18 June when the source was $I\sim 17$ 
using the 1.9 m SAAO telescope and grating spectrograph at a resolution of
about $7\AA$.

\section{Analysis}

	The proper motion can be determined using equation (\ref{eqn:mudef}) 
from measurements of the source crossing time, $t_*$, and 
the angular source radius, $\theta_*$.  The crossing time is extracted from
the light curve, while the source radius can be measured by two independent
methods, spectroscopic and photometric.

\subsection{Measurement of $t_*$}

	A static binary lens with a uniform surface-brightness
finite source generates a light curve characterized by nine parameters.
Three of these are $\te$, $t_0$, and $u_0(\geq 0)$, 
the Einstein radius crossing
time, the time of closest approach, 
and the impact parameter of a hypothetical
event in which the binary is replaced by a single point-lens  of the same
total mass placed
at the geometrical midpoint of the binary.  Two
other parameters are $I_s$ and $I_b$, the magnitudes of the lensed source
and of any unresolved (unlensed) light superposed on the source star, 
respectively.  
Three parameters specific to the binary character of the lens are
the mass ratio $q=M_2/M_1$ of the secondary to the primary ($0<q\leq 1$),
the projected binary separation $d$ in units of the (combined)
Einstein radius $\re$,
and the angle $\alpha$ $(0\leq\alpha<2\pi)$
between the binary-separation vector ($M_2$ to $M_1$)
and the proper motion of the source relative to the binary center of mass
(see Fig.\ \ref{fig:two}).  The direction of the relative motion between 
lens and source is chosen so
that the lens system is on the right hand side of the moving source.
  Finally, $\rho_*\equiv \theta_*/\theta_{\rm E}$
is the source radius in units of the angular Einstein radius.  

Once the full solution
is found, the crossing time is given by $t_* = \rho_* \te$.  In order
to understand the sources of uncertainty in this determination, however, it is
useful to write the equation as,
\begin{equation}
t_* = \Delta t \sin\phi,\qquad \Delta t \equiv \rho_* 
\te\csc\phi,\label{eqn:tstar}
\end{equation}
where $\phi$ is the angle at which the source crosses the caustic
(see inset to Fig.\ \ref{fig:two}).  The
quantity $(2\Delta t)$ is the time taken by the source to
transit the caustic.  If the caustic crossing is well monitored (as it is
in this case) then $\Delta t$ is determined to high
precision.  We find below that $\Delta t = 4.25$ hours.  
Most of the uncertainty in $t_*$ therefore comes
from the angle $\phi$, which is determined from the {\it global} fit of the
light curve.  If the data permit two or more discrete geometries, 
then $\sin\phi$ can take on substantially different values.

	Figure \ref{fig:one} 
shows data (nightly binned except at the peak) 
together with the light curves for our
models I and II.  Figure \ref{fig:two} shows the lens 
geometry, the resulting
caustic structure, and the trajectory of the source relative to this 
structure for model I, which has parameters
$\te = 108.4\,$days, 
$t_0 = 27.83$ May 1998 UT, 
$u_0 = 0.065$, 
$I_s=22.15$, 
$I_b=21.43$, 
$\alpha=348^\circ \hskip-2pt .2$, 
$q=0.29$, 
$d=0.59$, and
$\rho_*=0.00112$.
In particular, $t_*=\rho_* \te = 2.92\,$hours.
The source center passed over the caustic at UT 18.120 June.
The overall fit has $\chi^2=394$ for 164 degrees of freedom. 
The model II parameters are
$\te = 80.3\,$days, 
$t_0 = 10.68$ June 1998 UT, 
$u_0 = 0.036$, 
$I_s=21.81$, 
$I_b=21.89$, 
$\alpha=351^\circ \hskip-2pt .4$, 
$q=0.95$, 
$d=0.56$, and
$\rho_*=0.00112$, 
with  $t_* = 2.15\,$hours, and $\chi^2=442$.
Note that the values of $t_*$ are quite different in the two
models despite the fact that values of $\Delta t$ (4.26 hours and 4.23 hours)
are very similar.  The difference is due to the different angles at which
the source crosses the caustic, $\phi=43^\circ\hskip-2pt .2$ and
$30^\circ\hskip-2pt .6$ respectively.
The differences in $t_*$ and $I_s$ lead to a proper 
motion that is larger for model II by a factor 
of 
$1.59$.  The model II $\chi^2$ is $\sim 48$ higher
than for model I.  However, the fact
that model I has a better fit does not 
rule out model II, which gives an acceptable
fit to the data.  Hence, the difference in proper motions between the two 
models (which is much larger than the statistical error of each solution) 
gives a good indication of the systematic uncertainty.  It is possible that
this uncertainty will be removed as we acquire more data, or by combining our
data set with those of MACHO and EROS.  In particular, we note that model I
appears more consistent with the pre-first-caustic data that
MACHO put on their web site.  However, at present we must allow that
our proper motion could be off by a factor $\sim 1.6$.

We have also fit for $V_s$ and $V_b$ holding all seven other parameters fixed
at the values found for the $I$ band solutions.  For model I, we find
$[V,(V-I)] = (22.45,0.3)$, (21.75,0.3), and (21.30,0.3) for the source,
background, and total light respectively.  For model II, we find
(22.10,0.3), (21.80,-0.1), and (21.20,0.1).  Initially, the MACHO web page
estimated the total baseline values to be $V=21.8$, $V-R=0.1$.  After the
appearance of this {\it Letter} as a preprint, MACHO (A.\ Becker 1998
private communication) informed us that they had revised their estimate
to $V=21.4$.  In addition, OGLE (A.\ Udalski 1998 private 
communication) confirmed this value, finding $V=21.41\pm 0.23$.  Since our
fits contain no data from the baseline, these independent measurements of
the baseline flux serve as an important check on the viability of the models.


	Our determinations may be compared with those of the EROS collaboration
(Afonso et al.\ 1998) based on a single night of data covering the end of the
caustic crossing.  They found a lower limit $\Delta t > 3\,$hours and measured
the end of the caustic crossing at UT 7:08 $\pm0$:02 18 June 1998.  In our
model I, $\Delta t = 4.26$ hours, 
and the end of the caustic crossing is at UT 7:11 18 June 1998.

\subsection{Measurement of $\theta_*$}

	We estimate the angular radius of the source, $\theta_*$, from
the light curve.  The model I fit to the overall light curve yields an 
unlensed source magnitude $I=22.15$.  
We determine the color of the source, $V-I=0.31\pm0.02$, 
from its measured color near the caustic,  transforming our instrumental
magnitudes to standard Johnson/Cousins bands using a sample of stars in
our data set that were independently calibrated by OGLE (Udalski et al.\ 1998).
\ \ Note that while the color of the source at any particular time on the 
light curve
could be affected by blending, this effect is extremely small near the caustic
because the source is magnified by a factor $\sim 100$.  We assume an
extinction $A_V=0.22\pm 0.1$  (see below).
This implies $I_0 = 22.02$ and $(V-I)_0=0.22$.  At fixed
$(V-I)_0$ color, the effective temperature depends only very weakly on
metallicity (which we take to be [Fe/H]$=-0.75$).  We use the Yale Isochrones 
(Green, Demarque, \& King 1987) to
estimate $T_{\rm eff}\sim 7900\pm 300$~K.  Combining this with the 
dereddened flux,
and again using the Yale Isochrones, we find $\theta_*=0.089\,\mu$as.

	We now investigate the effects of various errors.  First an
extinction error of $\delta A_V$, alters $I_0$ by
$-\delta A_I\sim -0.6\delta A_V$ (Stanek 1996), leading to an 
incorrect estimate of the radius by 
$(\delta \theta_*/\theta_*)\sim (\ln 10/5)0.6\delta A_V$.  This would
be more than compensated, however, by the fact that the color would be
in error by $\delta E(V-I) \sim 0.4\delta A_V$.  This in turn would
lead to an incorrect estimate of the temperature and so of the 
$I$ band surface brightness, $\delta (2.5\log S) \sim 
\gamma \delta E(V-I)$, where
$\gamma\equiv
[(\partial \ln B_V /\partial T)/(\partial \ln B_I /\partial T)-1]^{-1}$, and
$B_V$ and $B_I$ are the Planck intensities at $0.55\,\mu$m and
$0.80\,\mu$m, respectively.  This causes an error in the radius of
$(\delta \theta_*/\theta_*)\sim -(\ln 10/5)0.4\gamma\delta A_V$.  Thus, the
net error is $(\delta \theta_*/\theta_*)\sim 
(\ln 10/5)(0.6-0.4\gamma) \delta A_V\sim -0.25\delta A_V$, for $T\sim 8000$ K.

	By interpolating the extinction map of Schlegel, Finkbeiner,
\& Davis (1988) across the SMC, we find a foreground extinction of $A_V=0.12$.
We adopt an internal extinction of $0.1\pm 0.1$ mag, and obtain 
$A_V=0.22\pm 0.1$.  From the discussion above, the 
extinction uncertainty yields an uncertainty in the radius of $\sim 3\%$.
The 300 K uncertainty in the temperature, which arises primarily from the
uncertain color calibration of the models (M.\ Pinsonneault 1998 private 
communication) generates a 5\% uncertainty in the radius.  Finally, the
$\sim 10\%$ uncertainty in the unmagnified flux of the source produces
another 5\% uncertainty in the radius, 
for a final estimate of $\theta_*=0.089\pm 0.007\,\mu$as.

	From the spectrum, we find that the source temperature is consistent
with the value $T=8000$ K that was estimated from
the color $(V-I)_0\sim 0.22$, i.e., roughly an A6 star.  
The flux $(V_0=22.24)$ is in reasonable agreement with what one would expect
from a metal-poor star at 8000~K $(M_V\sim 3.2)$ at the distance of the
SMC ($\sim 60\,\kpc$.)
Figure \ref{fig:three} shows the source spectrum.

\section{Results}

	Combining the measurements of $\theta_*$ and $t_*$ we find
\begin{equation}
\mu = {\theta_*\over t_*} = 1.26\pm 0.10\,\kms\,\kpc^{-1},\label{eqn:mueval}
\end{equation}
for model I,
where we have assumed that the error in $t_*$ is much smaller than in 
$\theta_*$.  As noted in the previous section however, model II
yields $\mu=2.00\,\kms\,\kpc$.  Hence, the difference between acceptable 
solutions is much larger than the error in each solution.
Figure \ref{fig:four} 
shows the distribution of proper motions expected from a standard
$\rho_*(r)\propto r^{-2}$ isothermal halo characterized by a rotation speed
$v_c=220\,\kms$, and assuming an SMC distance of 60 kpc 
(Barnes, Moffett, \& Gieren 1993).  Since the proper motion of the SMC is 
difficult to measure and not yet
tightly constrained (e.g. Kroupa \& Bastian 1997),
we show curves for 4 separate assumptions: three with an SMC 
transverse speed of $250\,\kms$ oriented at $0^\circ$, $90^\circ$, and
$180^\circ$ relative to the reflex motion of the Sun, and the fourth with
no transverse speed.  In all four cases we find that the fraction of events
with proper motions smaller than that given by equation (\ref{eqn:mueval})
is less than 0.1\% for model I.
For model II the fraction is less than 0.5\%.
This would seem to imply that the lens is not in the halo.

	The main alternative explanation for the event is that
the lens is in the SMC.  In this case, its transverse speed relative to
the source is $v\sim 75\,\kms$ for model I and $v\sim 120\,\kms$ for model II.
The measured one-dimensional dispersion
of the SMC carbon stars is $\sigma\sim 21\pm 2\,\kms$ 
(Hatzidimitriou et al.\ 1997).  
Figure \ref{fig:four}
shows the expected distribution of proper motions assuming that both the
lens and the source transverse velocity distributions are characterized by 
this speed.  That is, the SMC material is assumed to be characterized by a
single component (but see below).  The fraction of all events
with proper motions $\mu>1.26\,\kms\,
\kpc^{-1}$ is $\sim 9\%$.  The fraction with
$\mu>2.00\,\kms\,\kpc^{-1}$ is $< 0.1\%$.

	Another possibility is that we are seeing
material that has been tidally disrupted and is moving transversely at
75--120$\,\kms$.  There is substantial evidence for such rapidly moving
material in {\it radial velocity} studies of the SMC.  The extensive HI 
observations of McGee \& Newton (1981) show that the SMC has four principal 
and overlapping  structures at velocities $v_h = 114$, 133, 167 and 
192 $\kms$.  
McGee \& Newton (1981) state that of the 506 lines of sight 
examined, 97\% show evidence for multiple peaks.  Staveley-Smith et al.\ 
(1997) present much more detailed HI maps and argue that these data imply not
multiple components but rather a complexly structured and perhaps 
disintegrating SMC.
Zhao (1998) has modeled
the SMC as comoving tidal debris and predicted that such a structure should
generate about one microlensing event per year.
Torres \& Carranza (1987)
summarize evidence for 4 peaks at $v_h=105$, 140, 170, and 190 $\kms$
from their own HII data plus earlier results for
HI, planetary nebulae, supergiants,
emission regions, and CaII.  However, the picture that they present is much
more of correlations between radial velocity and position on the sky.  Of 
course, to produce the observed proper motion, there must be material at
two different velocities along the same line of sight.  We therefore believe
that it is plausible, but by no means proven, that there is material within
the SMC that could produce the proper motion indicated by model II.  
(As discussed above, the proper motion associated with model I can be 
explained from the internal dispersion of clumps within the SMC alone.)

	If the lens does lie in the SMC, then the Einstein radius is 
approximately $\re\sim (4G M \dls/c^2)^{1/2}$, where $\dls$ is the lens-source
separation.  The observed time scale is $\te=\re/v=108\,$days or $82\,$days
for the two solutions.  Hence,
the mass is
\begin{equation}
M \sim {v^2 \te^2 c^2\over 4 G \dls} \sim 
0.53\,M_\odot\biggl({\dls\over 5\,\kpc}
\biggr)^{-1},\label{eqn:massdist}
\end{equation}
for model I and a factor 1.4 times larger for model II.
Various studies have found that the SMC is extended along the line of sight.
Although the original estimates of 20 kpc or more by Mathewson, Ford, \& 
Visvanathan (1986) are probably too high, the depth may be as high as 10 kpc
(Welch et al.\ 1987; Martin, Maurice, \& Lequeux 1989;
Hatzidimitriou, Cannon, \& Hawkins 1993).  Equation (\ref{eqn:massdist})
implies that if the lens-source separation is comparable to this depth, then
the total mass of the lens is that of a typical low-mass binary. 
The projected separation of the binary is $d \re = b v \te\sim 2.7\,$AU
or $d\re\sim 3.1\,$AU, 
which are also quite reasonable for a low-mass binary.  The period
would then be $P\sim 6 (\dls/5\,\kpc)^{1/2}$ years, indicating that the effect
of internal binary motion during the event is small.

	In sum, MACHO-98-SMC-1 appears to be consistent
with lensing by a stellar 
binary in the SMC proper, and inconsistent with lensing by 
an object in the Galactic halo.

\begin{acknowledgements}

We thank the SAAO, CTIO, Canopus and Perth Observatories for generous awards 
of time to PLANET science.  
We thank David Gonzalez, Juan Espinoza, and Charles Bailyn who
obtained much of the CTIO-Yale 1 m data.
We thank Philip Keenan and Arne Sletteback for valuable discussion about
the spectral type of the source.
We thank B.\ Atwood, T.\ O'Brien, P.\ Byard, and the entire
staff of the OSU astronomical instrumentation lab for
building the instrument used at the CTIO-Yale 1-m telescope.
We give special thanks to the MACHO collaboration for making the public
alerts of this event and its probable impending
second caustic crossing, which made possible the observations reported here.
This work was supported by grants AST 97-27520 and AST 95-30619 from the NSF, 
by grant NAG5-7589 from NASA, by a grant from the Dutch ASTRON foundation, and
by a Marie Curie Fellowship from the European Union.  

\end{acknowledgements}

\newpage

\begin{figure}
\caption[junk]{\label{fig:two}
Geometry of model I.  The diamond shaped curve is the caustic, 
i.e., the region of formally infinite magnification.  The thick solid line
and two thinner parallel lines indicate the source trajectory and the
finite size of the source.  The tick marks are in units of the
Einstein radius crossing time, $\te=108.4\,$days.  The two components of
the binary are shown by {\it circles} whose relative sizes are proportional
to their masses.  Their projected separation in units of the Einstein
ring is labeled $d$.  The insert shows more clearly the finite size of the 
source as well as the angle $\phi$ at which the source intercepts the 
caustic.
}
\end{figure}

\begin{figure}
\caption[junk]{\label{fig:one}
Light curve of the PLANET data for MACHO-98-SMC-1.  Shown are the data from
the SAAO 1 m ({\it circles}), the CTIO 0.9 m ({\it squares}), the 
CTIO-Yale 1 m ({\it triangles}), and the Canopus 1 m ({\it asterisks}).  
The inset covers about 0.6 days, corresponding to less than one tick mark
on the main figure.
The data are binned by day (except near the
caustic) for clarity.  The light-curve fits were done without binning.
The two curves show model I ({\it bold}) and
model II ({\it solid}).  The two models have very similar
crossing times, $\Delta t\sim 4.25$ hours (which are determined from the
caustic crossing), but different angles at which the
lens crosses the caustic (see Fig \ref{fig:two}), 
$\phi=43^\circ\hskip-2pt .2$ versus
$\phi=30^\circ\hskip-2pt .6$ 
(which are determined from the overall light curve).  
The proper motion is determined from the combination $\mu = \theta_*/t_* =
\theta_*/(\Delta t\sin \phi)$ 
where $\theta_*$ is the angular size of the source.
}
\end{figure}

\begin{figure}
\caption[junk]{\label{fig:three}
PLANET spectrum of MACHO-SMC-98-1 taken near 18.0 June 1998 UT. 
The obvious 
Balmer line absorption indicates the source is an A-type star,
consistent with the spectral type estimated 
from the $V-I$ color of the source. 
The strength of weak CaII H and K absorption suggests 
the effective temperature of the star is close to 8000K, although 
the Balmer line absorption strengths seem to indicate a slightly
higher temperature. Note that the spectrograph slit was not well
aligned with the parallactic angle, so the overall spectral
energy distribution may be affected.
}
\end{figure}

\begin{figure}
\caption[junk]{\label{fig:four}
Distribution of expected proper motions from Galactic halo (right) and
non-tidal SMC (left) lenses.  The SMC model is based on a one-dimensional
dispersion of $21\,\kms$ for both the lenses and sources.
The normalizations are set arbitrarily to
unity at the peaks.  Four models are shown for the halo, three where 
the SMC is moving transversely at $250\,\kms$ at $0^\circ$ ({\it solid}), 
$90^\circ$ ({\it dashed}), and $180^\circ$ ({\it bold dashed})
relative to the reflex motion of the Sun, and one
where the SMC has no transverse motion ({\it bold}).  The boxes shows the
best fit and $1\,\sigma$ errors for the proper motion of the lens 
for model I (left) and model II (right).  The
central values lies below 99.9\% (99.5\%) of the halo distribution and 
above 91\% (99.9\%) of
the SMC distribution.  The observed value is inconsistent with the halo
hypothesis.  The lens and source could possess normal SMC kinematics 
or one of them could be part of SMC tidal debris.
}
\end{figure}

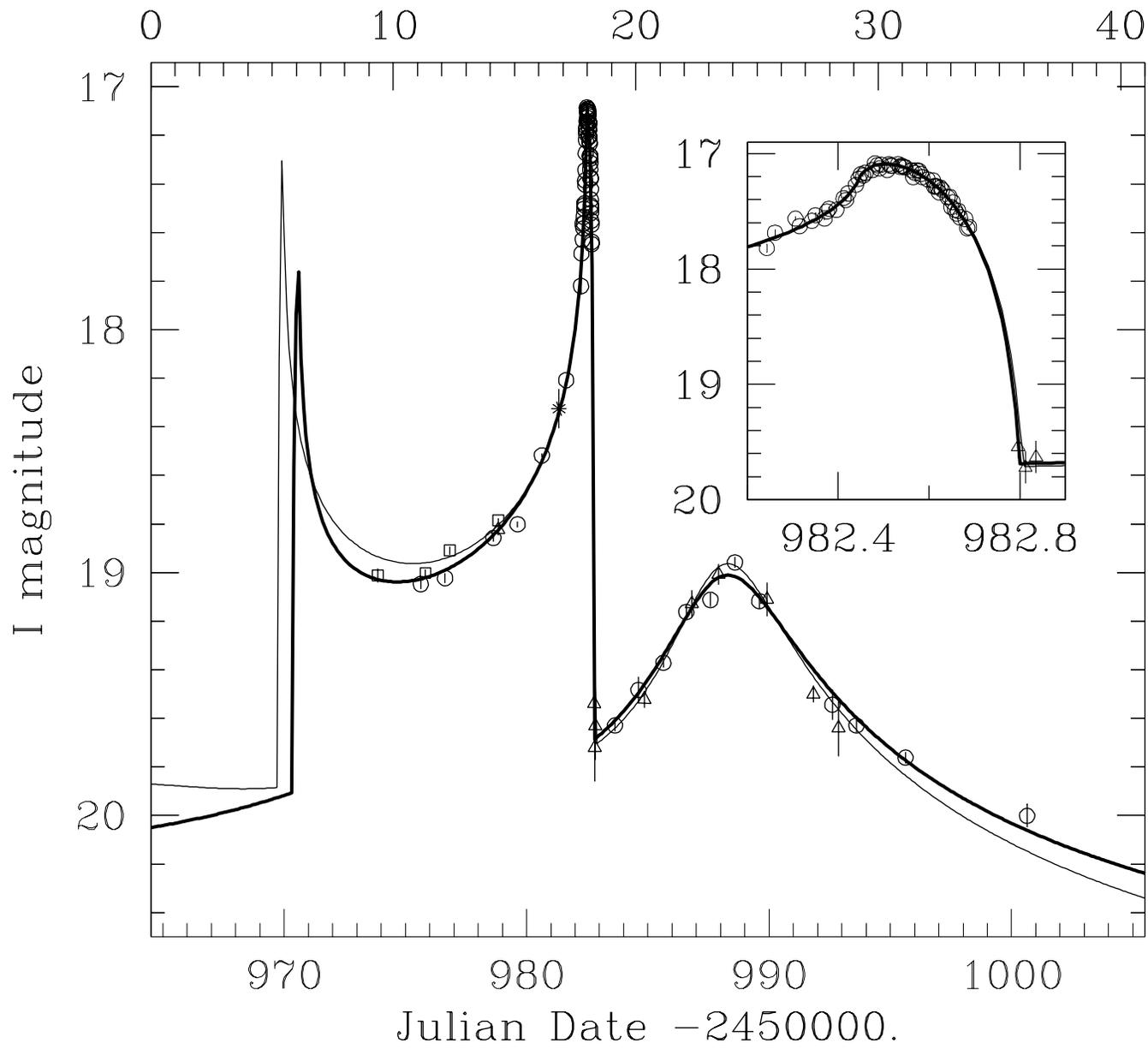
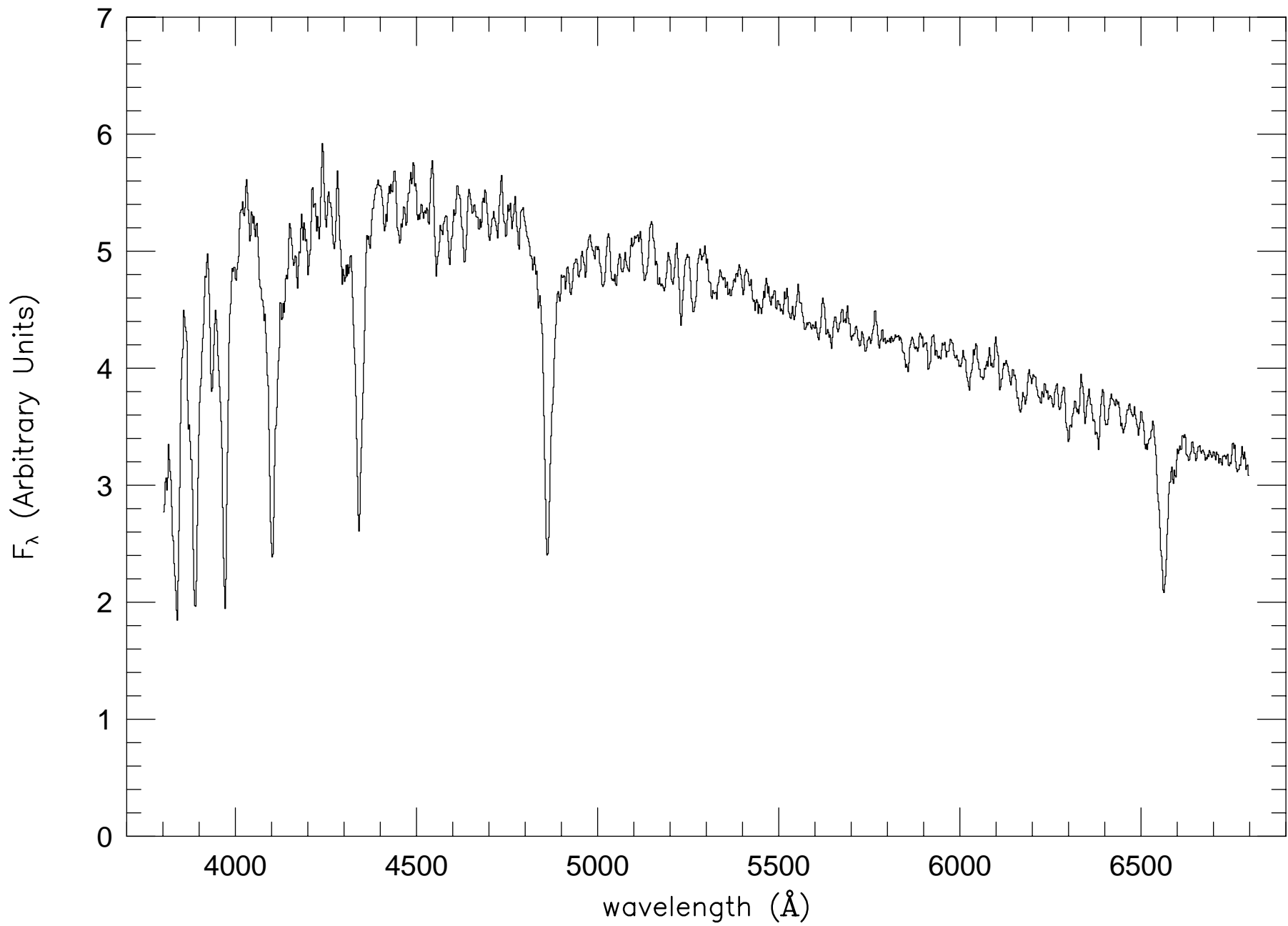
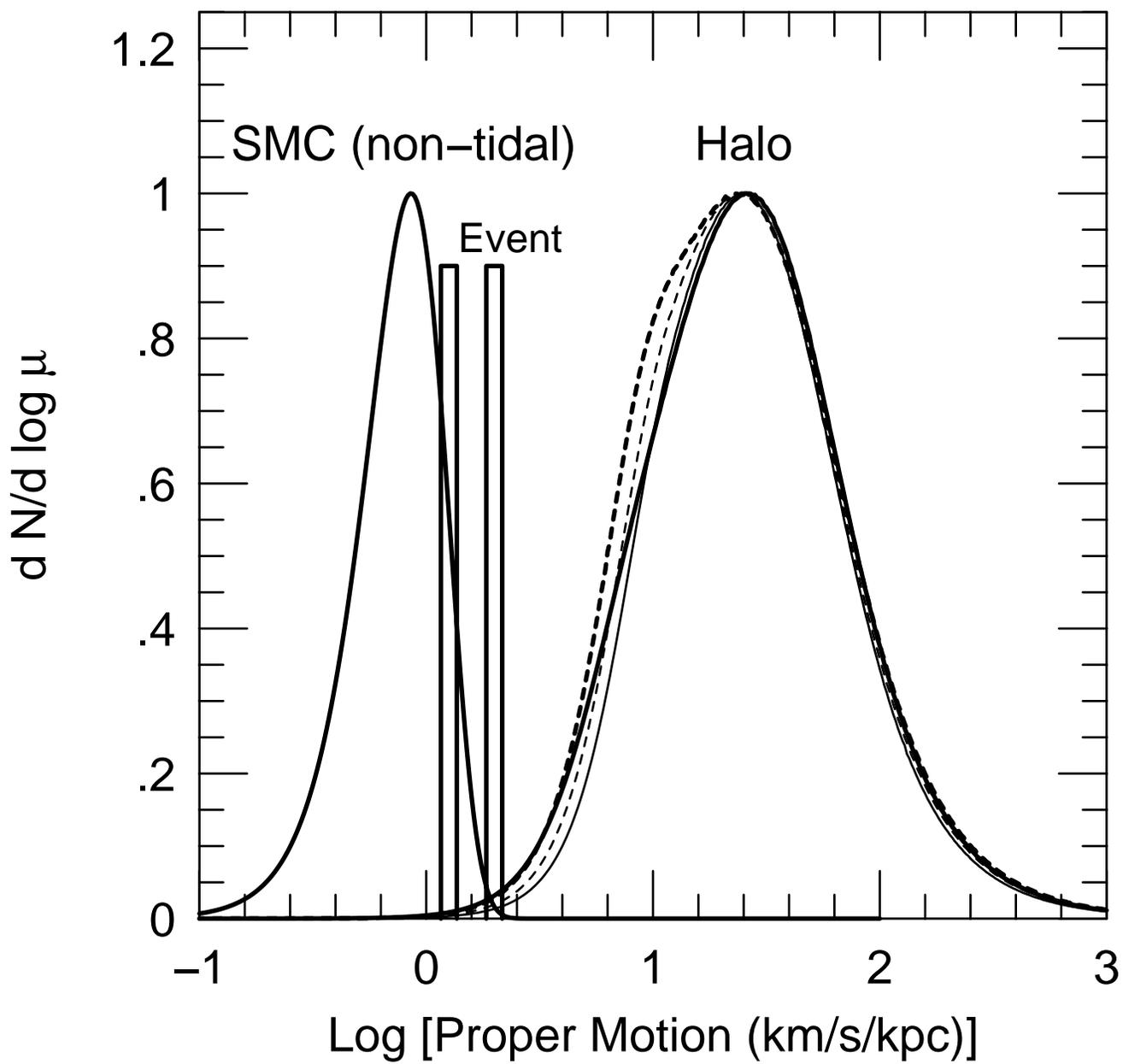
\end{document}